\documentclass{vienna-conf2019}
\usepackage{graphicx}
\usepackage{hyperref}
\usepackage[]{natbib}  
\usepackage{epstopdf}

\def\BibTeX{{\rm B\kern-.05em{\sc i\kern-.025em b}\kern-.08em
    T\kern-.1667em\lower.7ex\hbox{E}\kern-.125emX}}
\bibpunct{(}{)}{;}{a}{}{,}  

\begin{document}

\TitreGlobal{Stars and their variability observed from space}


\title{What physics is missing in theoretical models of high-mass stars: new insights from asteroseismology}

\runningtitle{New insights from massive star asteroseismology}

\author{D. M. Bowman}\address{Institute of Astronomy, KU Leuven, Celestijnenlaan 200D, B-3001 Leuven, Belgium}





\setcounter{page}{237}


\maketitle


\begin{abstract}
Asteroseismology of massive stars has recently begun a revolution thanks to high-precision time series photometry from space telescopes. This has allowed accurate and robust constraints on interior physical processes, such as mixing and rotation in the near-core region of stars, to be determined across different masses and ages. In this review, I discuss recent advances in our knowledge of massive star interiors made by means of gravity-mode asteroseismology, and highlight some new observational discoveries of variability in some of the most massive stars in our universe.
\end{abstract}



\section{Introduction}
The lives and energetic deaths of massive stars play a pivotal role in shaping the Universe \citep{Maeder_rotation_BOOK, Langer2012}. Massive stars are significant metal factories and provide energy and chemical feedback to the interstellar environment when they end their lives as a supernovae.  Hence, understanding how massive stars evolve is of paramount importance for the chemical and dynamical evolution of the host galaxy \citep{Bromm2009b, deRossi2010d}. Despite the importance of massive stars, the physics of their interiors is currently not well constrained, which in turn strongly impacts their post-main sequence evolution \citep{Ekstrom2012a, Chieffi2013}. Specifically, the interior mixing, rotation and angular momentum transport mechanisms inside massive stars are controlled by uncalibrated free parameters in evolution models \citep{Aerts2019b}. When combined with the significant effects of binarity and metallicity (e.g. \citealt{Sana2012b} and \citealt{Georgy2013c}), these represent large theoretical uncertainties in massive star evolution theory which need to be mitigated \citep{Martins2013c, Aerts2019b}.

A powerful method for probing stellar interiors is asteroseismology \citep{ASTERO_BOOK}, which uses stellar oscillations to probe the physics of stellar structure. Different types of pulsation modes can be excited within a massive star. Gravity modes are standing waves restored by buoyancy (i.e. gravity) and are extremely sensitive to the physics of near-core region in massive stars \citep{Miglio2008b}. For rotating stars, the Coriolis force is also a dominant restoring force, such that stars exhibit gravito-inertial modes which probe rotation in the near-core region \citep{Bouabid2013}. Massive stars can also pulsate in pressure modes, which probe the envelope and near-surface layers \citep{ASTERO_BOOK}. The successful application of asteroseismology requires long-term, continuous and high-precision photometric time series data to resolve individual pulsation mode frequencies, such that a quantitative comparison of observed and predicted pulsation modes frequencies reveals the physics that best represents the observed star. 

Since massive stars have convective cores and radiative envelopes during the main sequence, the physics of convection and convective-boundary mixing is crucial in determining their core masses and ultimate evolutionary fate \citep{Ekstrom2012a, Chieffi2013, Georgy2013c}. The mixing profile at the interface of the convective core and radiative envelope, and the mixing profile within the envelope itself directly impact the amount of hydrogen available for nuclear burning. With more internal mixing, a massive star experiences a longer main sequence lifetime and produces a larger helium core at the terminal age main sequence. The relative abundance of pulsating B stars compared to O stars means that the majority of constraints on massive star interiors currently come from $\beta$~Cep and slowly-pulsating B (SPB) stars \citep{Aerts2019b}. Together these two types of main sequence and post-main sequence pulsators span a wide range in mass between approximately 3 and 20~M$_{\odot}$. Hence these pulsators provide invaluable potential to constrain interior mixing and rotation in stars that span the boundary between intermediate- and high-mass stars – i.e. the boundary between stars that end their lives white dwarfs and those that explode as supernovae and become neutron stars or black holes \citep{Langer2012}.

\section{New insights of stellar interiors from asteroseismology}

Even though the pressure mode pulsations in $\beta$~Cep stars provide constraints on the interior properties of massive stars (see e.g. \citealt{Aerts2003d, Handler2004b,Handler2006a, Briquet2007e, Daszy2013b}), the focus of recent asteroseismic studies has been on gravity-mode pulsators (see \citealt{Aerts2019b} for a review). A powerful diagnostic in interpreting the oscillation spectrum of a star pulsating in gravity modes is its period spacing pattern, which is defined as the period differences of consecutive radial order ($n$) gravity modes of the same angular degree ($\ell$) and azimuthal order ($m$) as a function of the pulsation mode period. Under the asymptotic approximation, gravity modes in a non-rotating, chemically homogenous star are equally spaced in period, yet rotation and a chemical gradient left behind from a receding convective core introduce perturbations in the form of a “tilt” and “dips”, respectively \citep{Miglio2008b, Bouabid2013, VanReeth2016a}. Higher rotation rates induce a larger “tilt” with the gradient being negative for prograde modes and positive for retrograde modes. On the other hand, the “dips” caused by mode trapping are diminished with increased mixing \citep{Miglio2008b}. 

\begin{figure}[ht!]
\centering
\includegraphics[width=0.99\textwidth,clip]{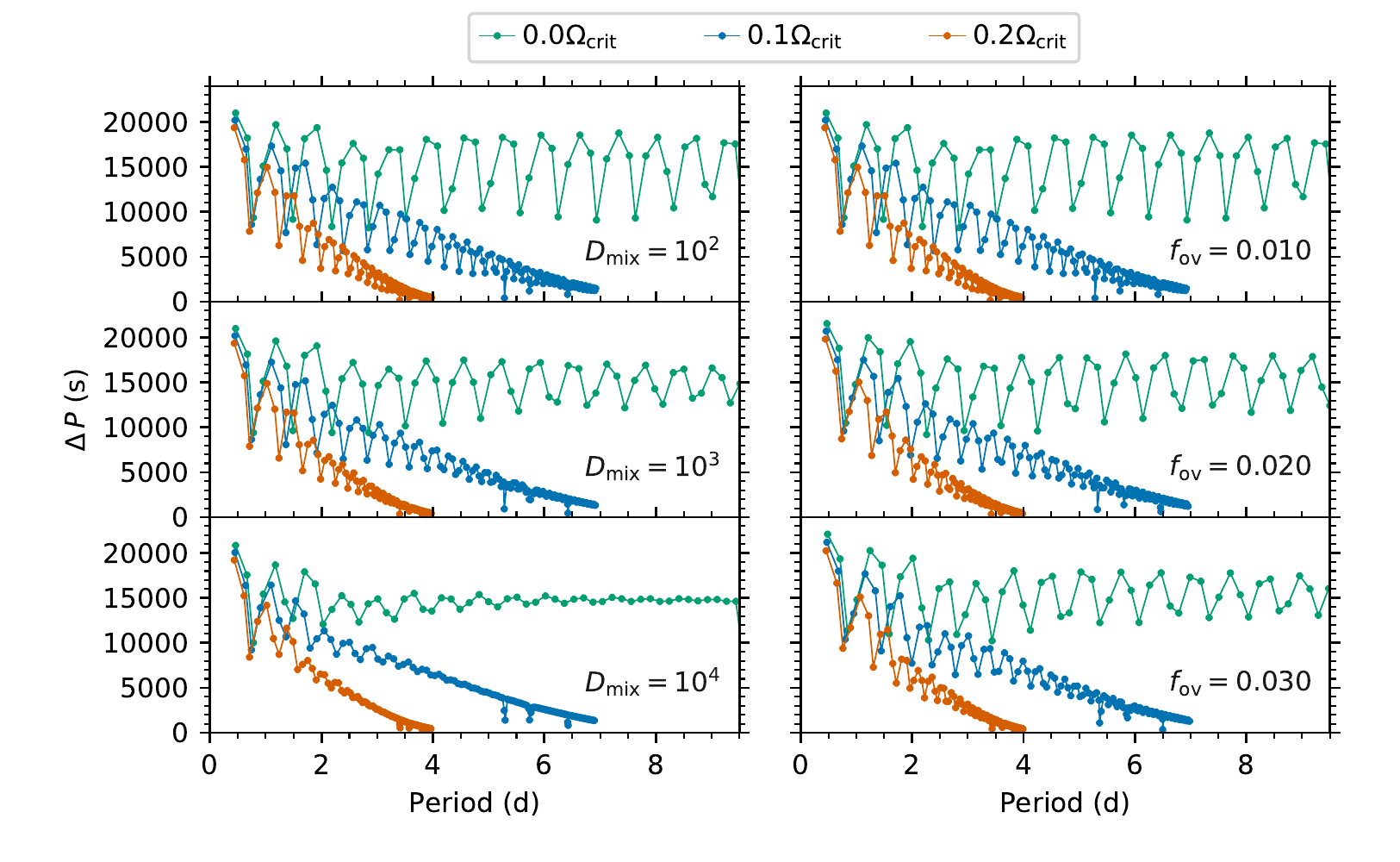}      
\caption{Theoretical gravity-mode period spacing patterns for prograde dipole modes of a 12~M$_{\odot}$ star about halfway through the main sequence (i.e. $X_{\rm c} = 0.4$). The left and right columns are for different envelope mixing ($D_{\rm mix}$) in cm$^{2}$\,s$^{-1}$, and exponential convective-boundary mixing ($f_{\rm ov}$) expressed in local pressure scale height, respectively, calculated using the {\sc MESA} stellar evolution code \citep{Paxton2011, Paxton2019}. For each panel, three rotation rates calculated using the Traditional Approximation for Rotation (TAR) using the {\sc GYRE} pulsation code \citep{Townsend2013b} are shown.}
\label{Bowman:fig1}
\end{figure}

An illustration of the effect of different amounts of interior mixing and rotation rates for a $12$~M$_{\odot}$ star about halfway through the main sequence is shown in Figure~\ref{Bowman:fig1}. Since a non-negligible rotation rate has a significant impact on stellar structure (see e.g. \citealt{Ouazzani2015a}), it also strongly affects the oscillation spectrum of a pulsating star \citep{Bouabid2013, VanReeth2016a, Aerts2019b}. Therefore, the effect of rotation is critical when calculating and interpreting the oscillation spectrum of a massive star \citep{Aerts2018b}, which includes slowly-rotating stars. The effect of (slow-to-moderate) rotation on gravity-mode frequencies, and also the corresponding period spacing pattern, is illustrated in Figure~\ref{Bowman:fig1}, which is calculated using the Traditional Approximation for Rotation (TAR) with the {\sc GYRE} pulsation code \citep{Townsend2013b,Townsend2018a}. Hence, the morphology of an observed gravity-mode period spacing pattern facilitates mode identification and offers a direct measurement of the near-core rotation and chemical mixing within a star.

\subsection{Interior mixing and rotation}

The high-precision space photometry assembled by the CoRoT \citep{Auvergne2009} and Kepler missions \citep{Borucki2010, Koch2010} heralded the birth of gravity-mode asteroseismology for massive stars. Although earlier studies primarily based on pressure modes extracted using ground-based photometry and/or spectroscopy provided valuable insights of massive star interiors, space photometry has opened the door to studying the deep interiors of many more stars across a wider parameter space in the Hertzsprung--Russell (HR) diagram than before. Gravity-mode asteroseismology of massive stars typically employs a data-driven approach, in which observations are used to calibrate models of stellar structure and evolution. In turn, this allows new insight of the physics currently missing within models.

The 4-yr Kepler photometric data set has unprecedented photometric precision and was the first data set to provide dozens of SPB stars to test theory. Amongst the first examples was the SPB star KIC~10526294. From a series of rotationally-split gravity modes in KIC~10526294, \citet{Papics2014} determined a near-core rotation period of approximately 188~d, which was in turn used by \citet{Triana2015} to compute a near-rigid interior rotation profile by means of an inversion. Later, the in-depth modelling of the 19 zonal dipole gravity-modes detected in KIC~10526294, allowed \citet{Moravveji2015b} to conclude that a non-zero amount of envelope mixing ($D_{\rm mix} \simeq 100$~cm$^2$\,s$^{-1}$) was needed to accurately explain the pulsation modes in this star. Furthermore, \citet{Moravveji2015b}, demonstrated that a non-negligible amount of extra mixing in the near-core region (also known as convective-core overshooting) was required to re-produce the oscillation spectrum of KIC~10526294 --- i.e. the best fitting models revealed $0.017 \leq f_{\rm ov} \leq 0.018$. Similar conclusions for needing extra mixing in a moderately rotating SPB star, KIC~7760680, were obtained by \citep{Moravveji2016b} with the best-fitting models yielding $D_{\rm mix} \simeq 10$~cm$^2$\,s$^{-1}$ and $f_{\rm ov} = 0.024 \pm 0.001$. At higher masses, moderate values of convective-core overshooting were also found for the 6-M$_{\odot}$ SPB star KIC~3240411 by \citep{Szewczuk2018a}, and a non-zero amount of overshooting in the magnetic, rapidly-rotating SPB star HD~43317 by \citet{Buysschaert2018c}. These pioneering asteroseismic studies demonstrated the power and importance of constraining the properties of convective cores using gravity modes in the era of space photometry, especially given the relatively large variance in measured overshooting values from only a small sample. In particular, observations clearly show that mixing, and hence core masses and main-sequence lifetimes of stars with convective cores are underestimated in current state-of-the-art models. 

Asteroseismology has since been applied to thousands of intermediate-mass stars observed by Kepler, covering masses between approximately 1 and 8~M$_{\odot}$, rotation rates up to 80\% of critical, and evolutionary stages from the main sequence through to the red giant branch. An important conclusion from such a large number of stars has been that current angular momentum transport theory is erroneous by more than an order of magnitude \citep{Aerts2019b}. The situation is less clear for massive stars owing to the much smaller sample size currently available, but significant progress has already been made in recent years because of space telescope data and gravity-mode asteroseismology.

In addition to demonstrating the need for larger convective cores in main sequence stars than currently predicted by models, asteroseismology also has the capability to ascertain the {\it shape} of the mixing profile within the convective-core overshooting region \citep{Pedersen2018a, Mombarg2019a}. Typical shapes available in models include the ``step" and ``exponential" overshooting prescription, but there is currently little consensus as to the correct amount and shape of the mixing profile for massive stars, the temperature gradient within the overshooting region (see e.g. \citealt{Michielsen2019a}), and how these may change as a function of mass and age. 

In addition to the need for convective-boundary mixing in massive stars, the origin of mixing within their radiative envelopes is also unconstrained within evolutionary models. Direct evidence for needing increased envelope mixing comes from enhanced surface nitrogen abundances in massive stars \citep{Hunter_I_2009a, Brott2011b}. Since nitrogen is a by-product of the CNO cycle of nuclear fusion in a massive star, an efficient mixing mechanism in the stellar envelope must bring it to the surface. Rotationally-induced mixing has been proposed as a possible mechanism \citep{Maeder2000a}, but it is currently unable to explain observed surface nitrogen abundances in slowly-rotating massive stars in the Milky Way and low-metallicity Large Magellanic Cloud (LMC) galaxies \citep{Hunter_I_2009a, Brott2011b}. Furthermore, there was no statistically-significant relationship between the observed rotation and surface nitrogen abundance in a sample of galactic massive stars studied by \citet{Aerts2014a}. In fact, the only robust correlation with surface nitrogen abundance in the sample was the dominant pulsation frequency \citep{Aerts2014a}, which suggests that pulsational mixing is important in massive stars \citep{Townsend2018a}. This clearly motivates ongoing work to constrain the origin, amount and shape of envelope mixing in massive stars using gravity-mode asteroseismology given the clear impact of interior mixing and rotation on spectroscopic surface abundances and stellar evolution.

\subsection{Diverse photometric variability in massive star photospheres}

With the successful delivery of TESS mission photometry \citep{Ricker2015}, we are now entering a new and exciting era in which asteroseismology can be applied to hundreds of pulsating massive stars \citep{Pedersen2019a}. The TESS mission is providing a large photometric data set, which also includes hundreds of massive stars in the low-metallicity environment of the Large Magellanic Cloud galaxy \citep{Bowman2019a}. The diversity of photometric variability in hundreds of massive stars -- i.e. those known to have spectral types O or B -- has already been demonstrated by \citet{Pedersen2019a}, a sample which also includes (eclipsing) binaries, pulsating stars and magnetic stars (see also, e.g., \citealt{Handler2019a} and \citealt{David-Uraz2019b}). 

\begin{figure}[ht!]
\centering
\includegraphics[width=0.6\textwidth,clip]{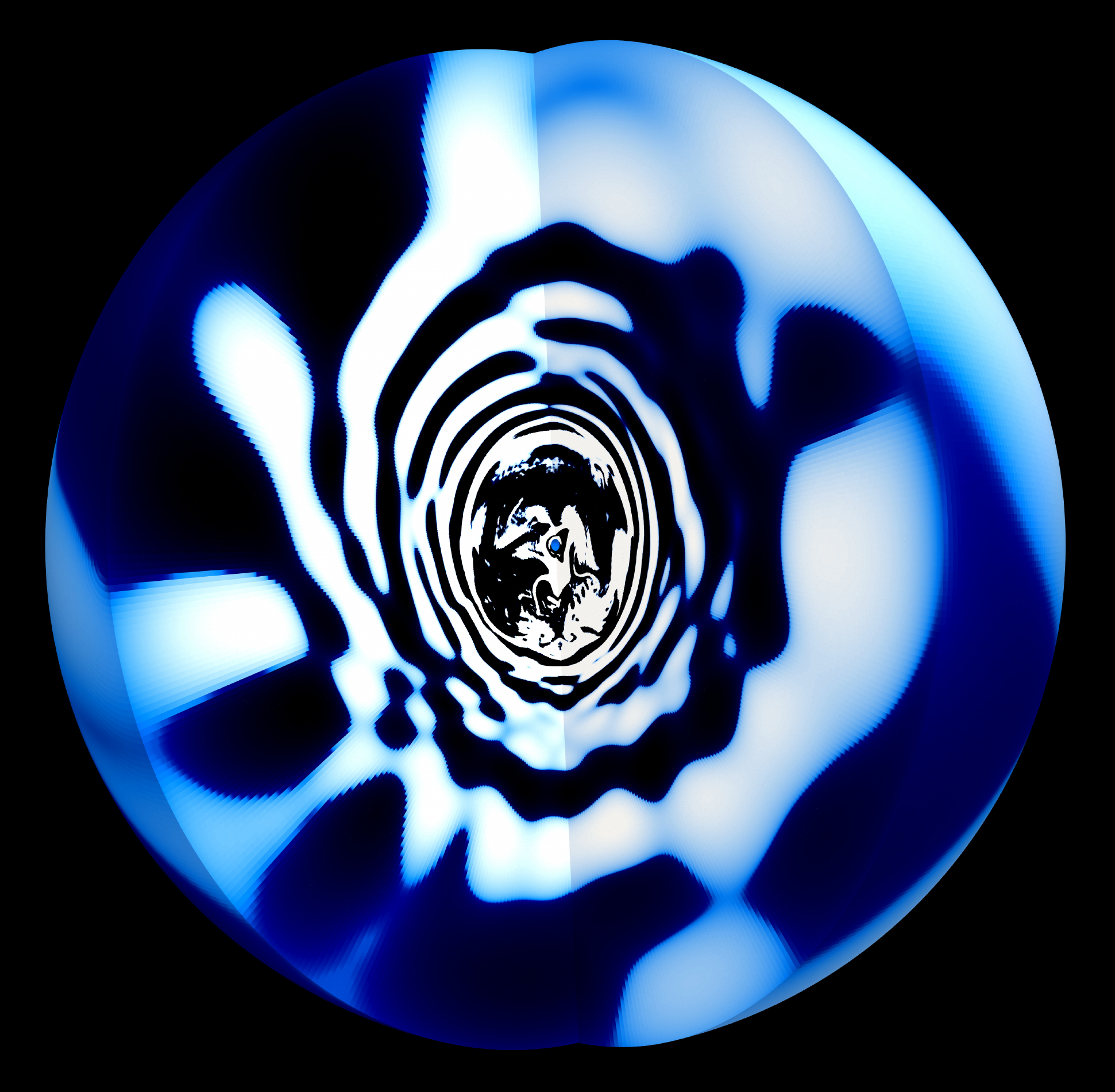}         
\caption{Snapshot of a 3D numerical simulation of internal gravity waves in a main-sequence massive star, with a white-blue colour scale for temperature fluctuations. Simulation courtesy of \citet{Edelmann2019a}.}
\label{Bowman:fig2}
\end{figure}

A recent discovery made using TESS mission photometry combined with data from the K2 mission \citep{Howell2014} was that the vast majority of stars with spectral types O or B have significant low-frequency variability in their light curves and amplitude spectra in addition to coherent pressure and/or gravity modes \citep{Bowman2019b}. Such stochastic variability is not predicted from pulsation excitation mechanisms commonly associated with coherent pressure and gravity modes in massive stars. However, convectively-driven gravity waves excited at the boundary of convective regions are predicted by 3D hydrodynamical simulations to produce low-frequency gravity waves and stochastic variability near the surface of a massive star \citep{Edelmann2019a}. An example of the temperature fluctuations caused by gravity waves driven by core convection within a massive star from a numerical simulation is shown in Figure~\ref{Bowman:fig2}.

\begin{figure}[ht!]
\centering
\includegraphics[width=0.49\textwidth,clip]{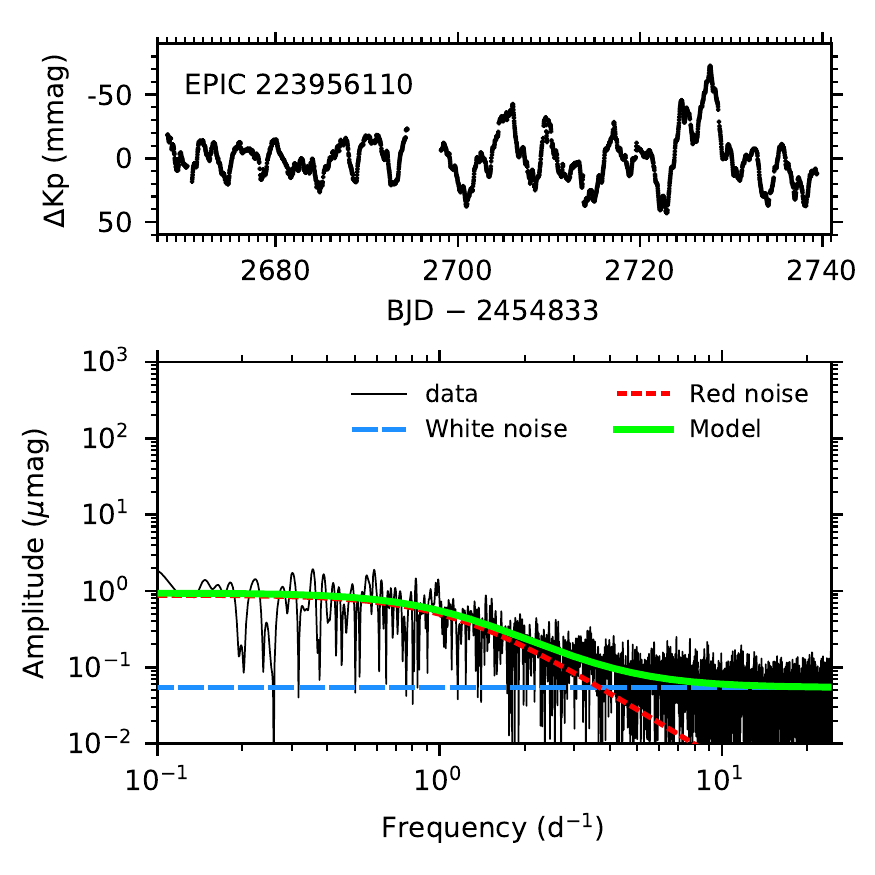}   
\includegraphics[width=0.49\textwidth,clip]{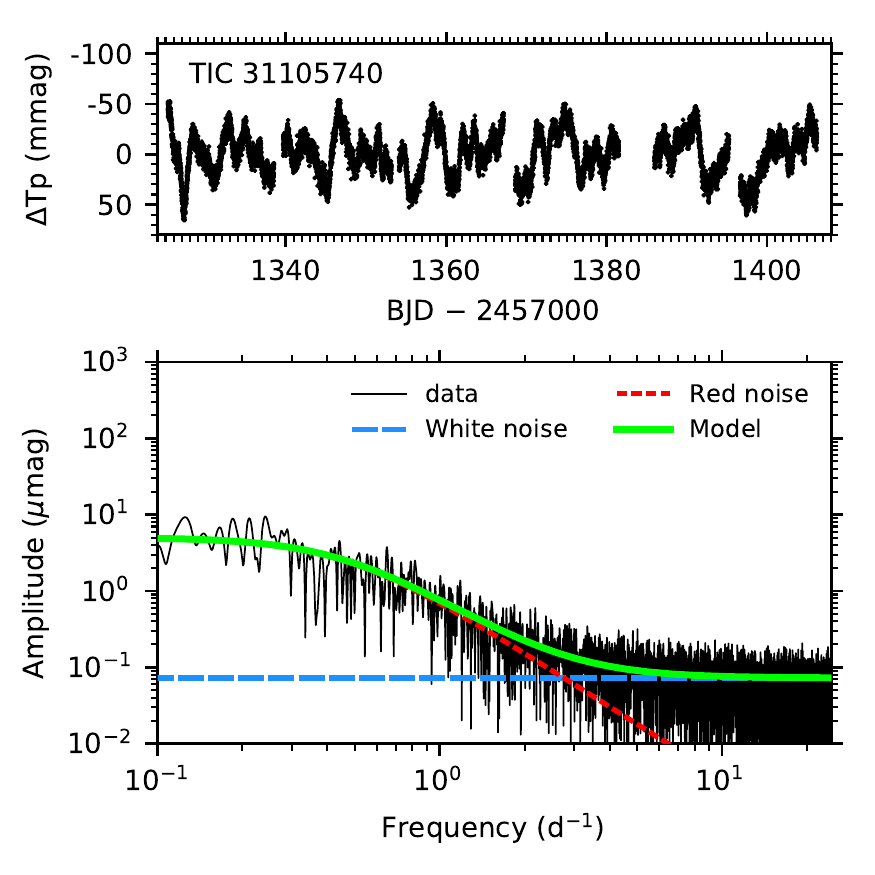}   
\caption{Observed light curves and amplitude spectra of two pulsating massive stars, which show significant low-frequency variability indicative of an entire spectrum of low-frequency gravity waves after coherent gravity and/or pressure modes have been removed by iterative pre-whitening \citep{Bowman2019b}. The left panel is the B0\,Ia(n) star EPIC~223956110 and the right panel is the B0.5\,Ia star TIC~31105740.}
\label{Bowman:fig3}
\end{figure}

The morphology of the low-frequency variability in more than 160 OB stars was found to be similar across a large range in stellar mass and age and, most importantly, the morphologies were also similar between metal-rich galactic and metal-poor LMC stars \citep{Bowman2019b}. The insensitivity of the low-frequency variability to the apparent metallicity of the host star and that evolutionary timescales predicted that most of the stars in the sample were likely to be in the main sequence phase of evolution led \citet{Bowman2019b} to conclude that the low-frequency variability was evidence of gravity waves excited by core convection. Two examples of massive stars with observed low-frequency variability after coherent pressure and gravity modes have been removed using iterative pre-whitening \citep{Bowman2019b} are shown in Figure~\ref{Bowman:fig3}, in which the left panel corresponds to the B0\,Ia(n) galactic star EPIC~223956110 observed by the K2 mission and the right panel corresponds to the B0.5\,Ia LMC star TIC~31105740 observed by the TESS mission.

At present, there are four excitation mechanisms known to trigger waves in OB stars: coherent pressure and/or gravity modes excited by a heat-engine mechanism \citep{Szewczuk2017a}, stochastic wave generation at the interface of the convective core and the radiative envelope \citep{Edelmann2019a}, stochastic wave generation by thin sub-surface convection zones \citep{Cantiello2009a, Lecoanet2019a}, and tidal excitation in binary systems \citep{Fuller2017c}. The identification of standing gravity waves within the observed low-frequency variability is essential to calibrate and constrain numerical simulations of convectively driven waves in terms of wave excitation, propagation and dissipation \citep{Edelmann2019a}, and ultimately facilitate asteroseismology in stars for which the heat-engine mechanism may not be the dominant excitation mechanism \citep{Bowman2019b}.

\section{Conclusions and future prospects}

Today, thanks to space missions including Kepler/K2 \citep{Borucki2010, Howell2014}, TESS \citep{Ricker2015} and BRITE-Constellation \citep{Weiss2014}, there is huge asteroseismic potential for massive stars. The long-term and high-photometric precision provided by space telescopes is unrivalled by ground-based telescopes, and the sample of massive stars is growing significantly larger thanks to the ongoing all-sky TESS mission. Crucially, TESS is also observing massive stars in different metallicity regimes because its southern continuous viewing zone includes the LMC galaxy, which will allow pulsation excitation models to be tested for metal-rich and metal-poor stars. The diverse variability of massive stars, which includes both coherent pulsators and those with low-frequency gravity waves \citep{Pedersen2019a, Bowman2019b}, enables asteroseismology for a sample of massive stars larger by two orders of magnitude compared to any that came before the TESS mission.

An important future goal of asteroseismology is to constrain the near-core and envelope mixing profiles, interior rotation profiles and angular momentum transport mechanisms inside massive stars, since insight of the physics in the near-core region of stars above approximately 8~M$_{\odot}$ is currently lacking compared to intermediate- and low-mass stars \citep{Aerts2019b}. In turn this will mitigate the currently large uncertainties in stellar evolution theory and lead to improved predictions of supernovae chemical yields and remnant masses. The future is bright from massive stars, and the goal to calibrate stellar structure and evolution models of massive stars using gravity-mode asteroseismology is now within reach.

\begin{acknowledgements}
DMB thanks the SOC and LOC for their organisation of a productive and enjoyable conference. Some of the research leading to these results has received funding from the European Research Council (ERC) under the European Unions Horizon 2020 research and innovation programme (grant agreement No. 670519: MAMSIE). The K2 and TESS data discussed here are obtainable from the Mikulski Archive for Space Telescopes (MAST) at the Space Telescope Science Institute (STScI), which is operated by the Association of Universities for Research in Astronomy, Inc., under NASA contract NAS5-26555. Funding for the K2 mission is provided by NASA’s Science Mission Directorate. Support to MAST for TESS data is provided by the NASA Office of Space Science via grant NAG5-7584 and by other grants and contracts. Funding for the TESS mission is provided by the NASA Explorer Program. The research presented here made use of the SIMBAD database, operated at CDS, Strasbourg, France; the SAO/NASA Astrophysics Data System; and the VizieR catalog access tool, CDS, Strasbourg, France. 
\end{acknowledgements}

\bibliographystyle{aa}  
\bibliography{Bowman_3k06} 

\end{document}